\documentclass[prl,aps,twocolumn, floatfix]{revtex4}
\usepackage{amssymb}
\usepackage{graphicx}
\usepackage{dcolumn}
\usepackage{bm,amsmath,verbatim}

\def\be{\begin{equation}}
\def\ee{\end{equation}}
\def\bea{\begin{eqnarray}}
\def\eea{\end{eqnarray}}
\def\bse{\begin{subequations}}
\def\ese{\end{subequations}}

\def\be{\begin{eqnarray}}
\def\ee{\end{eqnarray}}

\begin{document}
\title{Shiba impurity bound states as a probe of topological superconductivity and Fermion parity changing quantum phase transitions}
\author{Jay D. Sau}
\author{Eugene Demler}
\affiliation{ Department of Physics, Harvard University, Cambridge, MA, 02138, USA}
\date{\today}
\begin{abstract}
Spin-orbit coupled superconductors are potentially interesting candidates  
for realizing topological and potentially non-Abelian states with Majorana Fermions. We argue that time-reversal 
broken spin-orbit coupled superconductors generically can be characterized as having sub-gap states that are 
bound to localized non-magnetic impurities.  Such bound states, which are referred
 to as Shiba states, can be detected as sharp resonances in the tunneling 
spectrum of the spin-orbit coupled superconductors. The Shiba 
state resonance can be tuned using a gate-voltage or a magnetic field 
from being at the edge of the gap at zero magnetic fields to crossing zero energy 
when the Zeeman splitting is tuned into the topological superconducting regime. The zero-crossing signifies a 
Fermion parity changing first order quantum phase transition, which is characterized by a Pfaffian 
topological invariant.  
 These zero-crossings of the impurity level can be used to locally
 characterize the topological superconducting state from tunneling experiments.
\end{abstract}

\maketitle

\paragraph{\textbf{Introduction:}}
Majorana Fermions (MF) have been the subject of intense recent study,
both due to their fundamental interest as a new type of particle with non-Abelian statistics 
and their potential application in topological quantum computation
 (TQC)\cite{nayak_RevModPhys'08,Wilczek-3,Moore,Kitaev,Wilczek2}.
Topological superconductors \cite{schnyder} are promising candidates for 
the practical solid state realization of  MFs
 \cite{p_wave_sf,chuanwei,fu_prl'08,sau,long-PRB,alicea,roman,oreg}.
A simple topological superconducting (TS) 
system supporting MFs, which has attracted considerable experimental attention \cite{Wilczek-3}, consists of a
 spin-orbit coupled semiconductor in a magnetic field placed in contact with an ordinary superconductor
 \cite{sau,long-PRB,alicea,roman,oreg}. 
 It has been shown that the semiconductor proposal in  one-dimension, i.e. a semiconducting nanowire, can be driven into 
a TS phase  through the appropriate tuning of
 the semiconductor chemical potential or equivalently, the carrier density \cite{roman,oreg}. 
Such nanowires provide a realization of MFs at its ends in the same class as $p$-wave superconductors \cite{yakovenko1,kitaev}.
The TS state is predicted to be realized  in the nanowire 
when ever  the chemical potential $\mu$ with respect 
to the bottom of one of the electron sub-bands of the nanowire satisfies the constraint $|\mu|<\sqrt{\Delta^2+V_Z^2}$,
where $V_Z$ is the magnetic-field induced  Zeeman potential in the wire and $\Delta$ is the superconducting pairing potential 
induced in  the wire from contact with the $s$-wave superconductor~\cite{robustness}.
 The $s$-wave proximity effect on an InAs quantum wire, which also has a sizable SO coupling,
may have already been realized in experiments~\cite{doh}.
 Therefore, the TS phase in a semiconductor quantum wire may be one of the most experimentally 
promising approaches to realizing MFs with non-Abelian statistics.
 In fact, recent experimental results \cite{kouwenhoven_science} on the semiconducting wire system suggesting 
the existence of MFs in this system has created a great deal of interest in the 
physics communtiy \cite{Marchmeeting}.

\begin{figure}[ht]
\centering
\includegraphics[scale=0.31,angle=270]{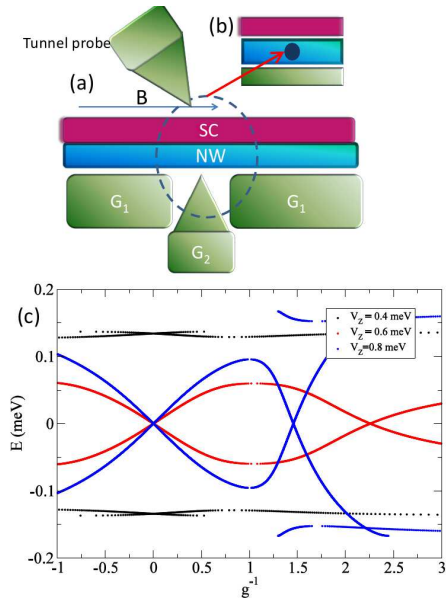}
\caption{(Color) (a) Shiba states in a nanowire (NW), which is in a magnetic field B and in proximity to a superconductor (SC)
 can be detected as a conductance peak from an STM tip or other tunnel probe near a gate-induced tunable potential $G_2$. 
(b) The gate potential produced by $G_2$ can be replaced by a local imupurity. 
 (c)  Shiba bound state energies as a function of inverse impurity strength or impurity transparency $g^{-1}$
 for various values of the Zeeman splitting $V_Z$ for a topological superconducting  nanowire (i.e. $V_Z=0.6$ meV (red curve),
 $V_Z=0.8$ meV(blue curve)) 
shows a characteristic zero-energy crossings, while the wire in the non-topological superconducting phase (i.e. $V_Z=0.2$ meV (black curve))
 shows weakly bound Shiba states. 
}\label{Fig2}
\end{figure}

While several schemes  such as tunneling and the fractional Josephson effect 
have been proposed to detect the presence of MFs \cite{kitaev,yakovenko,sau,roman,oreg},
 local probes that characterize the TS state of the wire are 
still missing. The need for such local characterization
 becomes specifically urgent because of the presence of disorder in realistic 
semiconducting wires. The absence of time-reversal symmetry leaves the proximity-induced 
superconducting gap of the nanowire susceptible to disorder \cite{tudor,piet,damle,anderson,potter_disorder}, which is predicted to lead to 
subgap states in the bulk of the wire \cite{roman_tudor,tudor}. Such disorder induced gapless states can  broaden the MFs
 into Majorana resonances and eliminate the fractional Josephson effect \cite{fu_kane,periodic}. 
Therefore, while probes of topological superconductivity such as the fractional Josephson effect, 
provide a true characterization of the topological degeneracy associated with MFs,
 they cannot distinguish if some part of the disordered wire is locally in an essentially  TS  
phase. Scanning tunneling microscopy (STM) of superconductors has provided one such route to experimentally characterizing,
in a local way, the superconducting state of superconductors such as the high-Tc cuprate superconductors \cite{balatsky}.
 In fact, the STM spectrum of impurities allows 
the characterization of not only the superconductor \cite{yazdani}, but also the effect of different impurities 
on the superconducting state \cite{apl}. Magnetic impurities in spin-singlet superconductors lead to Shiba states, 
which appear as sharp features in  tunneling spectra ~\cite{shiba}. Tuning the strength of the impurity has been 
predicted, in principle, to lead to a local quantum phase transition(QPT) ~\cite{sakurai,salkola}.  Therefore, it is natural 
to expect that studying the tunneling spectra around impurities in wires might allow one 
 to understand the TS character of each part of the wire in a local way.

\begin{figure}
\centering
\includegraphics[scale=0.3,angle=0]{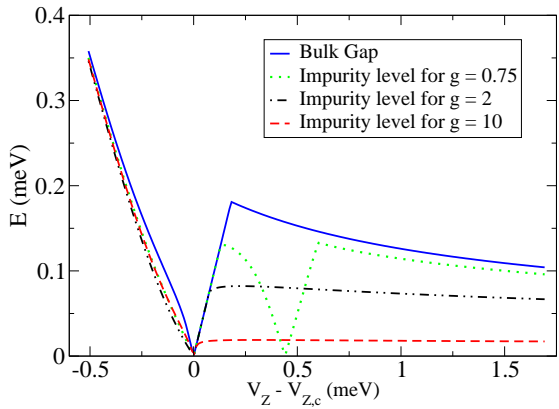}
\caption{(Color online) Impurity bound state energy and bulk gap as a function of Zeeman splitting for various values of 
impurity strength $g$. The values of $V_Z$ greater than the critical value $V_Z\sim 0.7$ meV, at which the bulk gap vanishes, 
support a topological superconducting phase in the nanowire.
 The Shiba bound state energies are seen to be deeply bound with energies significantly below 
the bulk gap only in the topological superconducting phase. 
Such Shiba bound states might contribute to near-zero bias conductance that is 
robust over a range of Zeeman potentials seen in recent experiments~\cite{kouwenhoven_science}.
 Here we have chosen $\mu=0.2$ meV, $\Delta=0.5$ meV and the spin-orbit coupling $\alpha=0.3$ corresponding 
to a spin-orbit energy $E_{SO}\approx 50\,\mu$eV~\cite{kouwenhoven}.}\label{Fig1}
\end{figure}

In this paper, we show that the three ingredients of spin-orbit coupling, Zeeman splitting and $s$-wave superconductivity 
required to realize the TS phase in nanowires are also the ingredients that realize 
sub-gap Shiba states bound to non-magnetic impurities.  Such Shiba states in semiconductor nanowires in proximity to a 
superconductor can be probed by a STM or tunnel probe arrangement shown in Fig.~\ref{Fig2}(a,b).
For static impurities, we find that the energy associated 
with the Shiba-state can be tuned to cross zero as seen in Fig.~\ref{Fig2}(c), providing a realization of the previously 
predicted Fermion-parity changing transition \cite{salkola} in a scenario where it might be easier to tune the 
impurity strength. Such a zero-crossing will be shown to be associated with a 
local version of the Pfaffian topological invariant that may be used to characterize the TS phase.
By a numerical study of models for the semiconducting nanowires, we find (see Fig.~\ref{Fig1}) that such strongly bound Shiba states only occur in 
the TS phase of the nanowires. The energy of these sub-gap states in the TS phase are found to be independent of the 
Zeeman energy and provide a possible alternative explanation to the recently observed zero-bias conduction peaks~\cite{Marchmeeting}.
 Such sub-gap states may be viewed as a way to destroy 
the TS gap, which is complimentary to the Born approximation approach applicable to short-scale disorder such as surface roughness of the 
nanowire. 
\paragraph{\textbf{Local impurity bound states in superconductors:}}
To understand the spectrum associated with impurities, we consider the Green function
 $G_{\tau\sigma;\tau'\sigma'}(\bm r;\bm r';E)$ of the superconducting nanowire in the Nambu spinor notation. 
Here $\bm r,\bm r'$ represent spatial coordinates on the nanowire, $\sigma,\sigma'$ refer to the spin 
indices of the electron and $\tau,\tau'=0,1$ represent the particle-hole index, which is needed to describe both 
the normal and anomalous parts of the Green function of the superconducting nanowire. The Green function $G$ matrix for 
an impurity in a superconductor can be calculated using the Dyson equation 
\begin{align}
&G(\bm r\bm r';E)=G_0(\bm r\bm r';E)\nonumber\\
&+\int d\bm r_1 G^{(0)}(\bm r\bm r_1;E)V(\bm r_1)\tau_z G(\bm r_1\bm r';E)\label{dyson}
\end{align} 
where $V(\bm r)\propto g\delta(\bm r)$ is the localized 
 impurity potential and $G^{(0)}_{\tau\sigma;\tau'\sigma'}(\bm r;\bm r';E)$
is the Green function of a clean superconducting nanowire with a Bogoliubov de-Gennes (BdG) Hamiltonian 
\begin{equation}
H_0=(-\partial_x^2-\mu+i\alpha\sigma_y\partial_x)\tau_z+V_Z\sigma_z+\Delta\tau_x.\label{HBdg}
\end{equation}
In the case of a single-channel wire, $\mu$ represents the chemical potential, $\alpha$ is the strength 
of Rashba spin-orbit coupling, $V_Z$ the magnetic field induced Zeeman 
potential and $\Delta$ represents the proximity induced superconducting pairing potential ~\cite{sau,roman}. 
The impurity strength $g$ is inversely related to the transparency $Z$ of the impurity potential. The matrices 
$\sigma_z$, $\tau_x$, and $\tau_z$ are Pauli matrices associated with the indices $\sigma$ and $\tau$ respectively. The  
BdG Hamiltonian $H_0$ can also be used to represent multi-band wires \cite{akhmerov,roman_tudor,potter} 
if $\mu$, $V_Z$, $\Delta$ and $g$ are taken to be matrices 
indexed by the channel index. For local impurity potentials, the energy levels of bound states are determined from the
 $\bm r=\bm r'=0$ part of the Dyson equation Eq.~\ref{dyson}, which is written as 
 \begin{align}
&G(00;E)=(1-g G^{(0)}(00;E)\tau_z)^{-1} G^{(0)}(00;E)\label{dyson1}.
\end{align} 
The Shiba bound state appears as a pole in the Green function $G$, which corresponds to a zero of the matrix
\begin{equation}
 Det[g^{-1}- G^{(0)}(00;E)\tau_z]=0.\label{chareq}
\end{equation} 

Consistent with Anderson's theorem \cite{anderson,potter}, the above equation describing sub-gap states bound to non-magnetic impurities, is 
found to have no solutions in the absence of a Zeeman potential i.e. for $V_Z=0$.
The absence of Shiba states continues to hold, even in the presence of a spatially 
uniform Zeeman potential $V_Z\neq 0$, if the spin-orbit coupling $\alpha$ vanishes, since  
the Zeeman splitting does not affect the wave-functions of the BdG Hamiltonian. 
Therefore, non-magnetic impurities can lead to localized Shiba bound states in wires only in the presence 
of a combination of Zeeman splitting and spin-orbit coupling, which are precisely the conditions to 
realize a TS phase. 

In contrast, for finite spin-orbit coupling $\alpha$, Zeeman potential $V_Z$ and superconducting pairing potential $\Delta$,
 numerical solutions of Eq.~\ref{chareq} plotted in Fig.~\ref{Fig2} show sub-gap Shiba states bound to even non-magnetic impurities. 
 While bound Shiba states are found to exist in the entire range of Zeeman potential $V_Z>0$, one observes that 
the Shiba states occur deep inside the bulk gap of the nanowire, which is also plotted in Fig.~\ref{Fig1}, only on the 
TS phase of the nanowire, i.e. $V_Z>V_{Z,c}=0.5$ meV. The TS state of the nanowire 
 can be seen identified in Fig.~\ref{Fig1} by the vanishing of the bulk gap in the wire 
at $V_Z=V_{Z,c}$. Thus, the existence of deeply bound Shiba states may be considered as suggestive of the 
nanowire being in the TS phase.

\paragraph{\textbf{Shiba states in  the TS phase:}}
While the detection of a deeply bound Shiba state might suggest the nanowire being in a TS phase, it cannot be 
taken as a confirmatory test.
Here we show that the evolution of the Shiba bound state energy as a function of the strength of the 
impurity $g$ (or magnetic field or density; see discussion below) may be used as a precise characterization of the TS phase.
 This is seen from the numerical results 
for the Shiba bound state energy in the TS and NTS phases of 
the nanowire, which are plotted in Fig.~\ref{Fig2}. The Shiba bound state energy in the TS phase in Fig.~\ref{Fig2} shows a characteristic pair of 
crossings of $E=0$ at both zero and non-zero values of the impurity transparency $g^{-1}$.

\paragraph{\textbf{Parity-changing QPT:}}
The crossing of the Shiba bound state at $E=0$ at  vanishing transparency $g^{-1}\sim 0$  is characteristic of the TS phase. This 
follows from the fact that an impurity with vanishing tranparency $g^{-1}=0$ splits the nanowire into two nanowires with 
a pair of zero-energy MFs at the impurity. The pair of zero-energy MF split linearly by tunneling across the 
as the impurity is tuned away from vanishing transparency. The zero-crossing of the energy of a Shiba-state is accompanied by a 
change of the Fermion-parity of the ground state. This is because the positive and negative energy eigenvalues, $\epsilon_1(g^{-1})>0$ and
 $-\epsilon_1(g^{-1})<0$ of the BdG 
Hamiltonian refer to the different occupancies of the associated Fermionic state $1$. The ground state of the system at $g_1^{-1}$ 
corresponds to the 
positive energy eigenvalue being unoccupied.  When the impurity strength $g^{-1}$ is tuned from $g_1^{-1}$ to $g_2^{-1}$, so that 
the positive and negative energy states cross zero and are interchanged i.e. $\epsilon_1(g_2^{-1})<0$ and  $-\epsilon_1(g_2^{-1})>0$, 
the system reaches a state where the positive energy state $1$ is occupied. Such a state with a positive energy eigenvalue occupied 
is an excited state of the Hamiltonian with $g_2^{-1}$. However, 
 this excited state has the same Fermion parity as the ground state at $g_1^{-1}$, i.e. before the energy crossing.
Since the excited state obtained by tuning the impurity to $g_2^{-1}$ is related to the ground state at $g_2^{-1}$ 
by adding a Fermion, the ground state at $g_2^{-1}$ must have a different Fermion parity from the 
ground state at $g_1^{-1}$. Therefore, the zero-energy crossing of the Shiba state is associated 
with a QPT where the ground state of the system changes its Fermion parity. This is analogous to the QPT
 proposed for tunable magnetic impurities in conventional superconductors ~\cite{salkola}. The spin-orbit coupled nanowire 
provides a realization of this interesting transition using a non-magnetic impurity, which can be tuned by a local gate voltage.

The Fermion parity change associated with zero-energy crossings of Shiba states at vanishing impurity transparency $g^{-1}\sim 0$
in the TS phase requires the existence of an odd number of zero-energy crossings at finite impurity transparency $g^{-1}\neq 0$. 
This follows from the fact that the total number of zero-energy crossings going from $g^{-1}=-\infty$ to $g^{-1}=\infty$ must be 
even, since both these points are associated with the ground state Hamiltonian of the nanowire with no impurity i.e. at $g\sim 0$. 
Thus, the TS phase is characterized by the Shiba bound state energy crossing zero an odd number of times as the limit of infinite 
impurity strength $g\rightarrow\infty$ is approached from at least one of the sides of either strong repulsive impurities or 
strong attractive impurities.

\paragraph{\textbf{Local Pfaffian topological invariants associated with impurities:}}  
To strengthen the argument that the zero-energy crossing is a topological QPT,
 , we compute the fermion 
parity using the Pfaffian topological 
invariant~\cite{kitaev}, which can be written in terms of the BdG Hamiltonian~\cite{parag}.
 The Pfaffian topological invariant for a particle-hole symmetric BdG Hamiltonian $H_{BdG}$ 
can be written in terms of the particle-hole matrix $\Lambda=\sigma_y\tau_y$ as $Q(H_{BdG})=\textrm{sgn}\left(Pf[H_{BdG}\Lambda]\right)$.
The particle-hole matrix $\Lambda$ is defined so that the particle-hole symmetry of $H_{BdG}$ can be written
 as $H_{BdG}\Lambda=-\Lambda H_{BdG}^*$, which is equivalent to the condition that $H_{BdG}\Lambda$ is
 anti-symmetric, which in turn allows the definition of the Pfaffian 
$Pf[H_{BdG}\Lambda]$.
Defining the QPT of the impurity problem requires a local definition of the Pfaffian invariant.
 To obtain such a local definition  
we note that zero-energy crossings at an impurity also imply zero-energy crossings of the Green function 
$U=G^{(0)}(0,0;E=0)-g^{-1}\tau_z$ (see Eq.~\ref{chareq}). Since we are considering an impurity in a gapped superconductor $G^{(0)}$ and 
the relevant matrix for determining zero-modes $U=G^{(0)}(0,0;E=0)-g^{-1}\tau_z$ is both Hermitean and particle-hole symmetric.
 Therefore the local Pfaffian invariant is defined as  
\begin{equation}
Q(g^{-1})=\textrm{sgn}\left(Pf[(G^{(0)}(0,0;E=0)-g^{-1}\tau_z)\Lambda]\right).\label{topinv}
\end{equation}
The above Pfaffian invariant can only change sign when the determinant of $U=G^{(0)}(0,0;E=0)-g^{-1}\tau_z$ vanishes, which corresponds to 
a solution of Eq.~\ref{chareq} at $E=0$ and therefore a zero-crossing of Shiba bound state energies. The Fermion parity changing transition of the ground 
state as a function of impurity strength $g$ is characterized by the topological invariant $Q(g^{-1})$ changing sign. Therefore, 
the local Fermion parity at the impurity can be calculated directly from computing the local Fermion-parity $Q(g^{-1})$ without 
computing the Shiba bound states. 

\begin{figure}
\centering
\includegraphics[scale=0.3,angle=0]{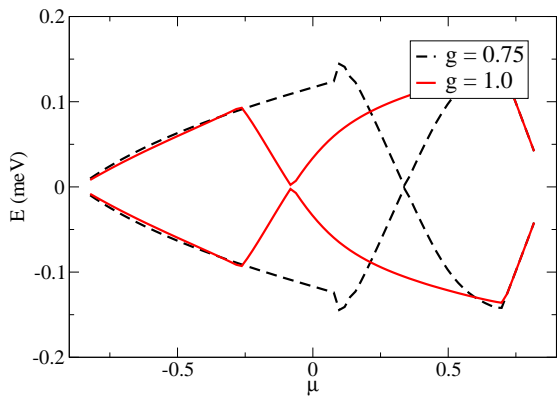}
\caption{(Color online) The evolution of the Shiba bound state energy with the local  chemical potential show an odd number of crossings
in the topological superconducting phase of the nanowire.
 The chemical potential in the neighborhood of an impurity in a nanowire can be varied, in principle, 
by a local gate voltage. The zero-energy crossings occur at different values of the local chemical potential for different values of the  
impurity transparency $g^{-1}$.}\label{Fig3}
\end{figure}

Even though the strength of an electrostatically induced impurity in a wire can in principle be controlled by a gate voltage in a 
simpler way than the tunable magnetic impurity required for $s$-wave superconductors \cite{salkola}, tunable non-magnetic impurities  
also might be difficult to obtain. An alternative approach to locally characterizing a TS phase 
is to study the evolution of the Shiba bound state energy as 
one tunes the chemical potential or the magnetic field towards the gap-closing topological phase transition. It is reasonable to 
expect a sufficiently strong impurity behaves has an asymptotically stronger effect as one approaches the phase transition. 
The evolution of the Shiba states across the phase transition for different local impurity strengths, which are 
plotted in Fig.~\ref{Fig1}(dotted green) and ~\ref{Fig3}, shows that this is indeed the case.
 Therefore, the TS phase of the nanowire can also be characterized
by the presence of an odd number of zero-energy crossings of the Shiba bound state energy as the applied chemical potential is 
locally tuned across a QPT.

\paragraph{\textbf{Conclusion:}}
Spin-orbit coupled coupled nanowires together with Zeeman splitting and $s$-wave superconductivity,  
which can be used to realize one-dimensional TS nanowires, are also precisely the ingredients needed to realize 
sub-gap Shiba states bound to non-magnetic impurities. Similar to magnetic impurities $s$-wave superconductors ~\cite{yazdani}, 
Shiba bound states can also be used to characterize the TS phase of the nanowire. Specifically, only the TS phase 
of the nanowire is found to support deeply bound Shiba states, whose energies are only weakly Zeeman field dependent. 
Such low-energy Shiba states found only in the TS phase provides a possible alternative explanation to the recent 
observations of zero-bias conductance peak, which was interpreted as a signature for the existance of MFs ~\cite{kouwenhoven_science}.
In the TS phase of the nanowire that the Shiba bound state energy is found to cross zero-energy an odd number of times 
as one tunes the strength of the impurity. Such zero-crossings of the Shiba bound state energy, besides providing 
a local characterization of the TS phase of the nanowire, are also intrinsically interesting since they are 
associated with a Fermion parity changing QPT of the nanowire ground state ~\cite{salkola}. For systems where the 
impurity strength is difficult to tune, we find that similar zero-energy crossings of the Shiba bound state 
energies for impurities in nanowires in the TS phase can also be obtained by gate tuning the chemical potential. 
Therefore, the TS phase of spin-orbit coupled semiconducting nanowires can be locally characterized by studying the 
evolution of Shiba bound state energies, which appear as sub-gap resonances in the tunneling spectrum into the wire.
Since most of our results follow from the scattering equation (Eq.~\ref{chareq}) from a local impurity, we expect our conclusions to apply to 
multi-channel nanowires as well as impurities in two-dimensional TS systems. 
Such local characterizations of the TS phase of the nanowire are particularly useful,
 since only parts of wires can be expected to enter the TS phase in 
disordered semiconducting wire. 

We acknowledge valuable discussions with Anton Akhmerov, Shou-Cheng Zhang and Ali Yazdani in the course of this work.
JS would also like to thank the Harvard Quantum Optics Center for support. ED would  
like to acknowledge the support from, Harvard-MIT CUA, NSF Grant No. DMR-07-05472.

\end{document}